\documentclass[twocolumn,trackchanges]{aastex701}

\usepackage{amsmath}
\usepackage{graphicx}
\usepackage{natbib}
\usepackage{color}
\usepackage{hyperref}
\usepackage{longtable}
\usepackage{rotating}
\usepackage{xcolor}
\usepackage{ulem}
\usepackage{lineno}

\linenumbers
\modulolinenumbers[10]

\begin{document}

\title{Galaxy mergers classification using CNNs trained on S{\'e}rsic models, residuals and raw images}

\author[orcid=0009-0000-2454-3788]{D. M. Chudy}
\affiliation{Astronomical Observatory of the Jagiellonian University, Faculty of Physics, Astronomy and Applied Computer Science, ul. Orla 171, 30-244 Kraków, Poland}
\email{dawid.chudy@doctoral.uj.edu.pl}

\author[orcid=0000-0002-7300-2213]{W. J. Pearson}
\affiliation{National Centre for Nuclear Research, Pasteura 7, 02-093 Warszawa, Poland}
\email{william.pearson@ncbj.gov.pl}

\author[orcid=0000-0001-7886-4915]{L. E. Suelves}
\affiliation{National Centre for Nuclear Research, Pasteura 7, 02-093 Warszawa, Poland}
\affiliation{Tartu Observatory, University of Tartu, Observatooriumi 1, Tõravere 61602, Estonia}
\email{luis.eduardo.suelves@ut.ee}

\author[orcid=0000-0001-8702-7019]{B. Margalef-Bentabol}
\affiliation{SRON Netherlands Institute for Space Research, Landleven 12, 9747 AD Groningen, The Netherlands}
\email{B.Margalef.Bentabol@sron.nl}

\author[orcid=0000-0003-3358-0665]{A. Pollo}
\affiliation{National Centre for Nuclear Research, Pasteura 7, 02-093 Warszawa, Poland}
\affiliation{Astronomical Observatory of the Jagiellonian University, Faculty of Physics, Astronomy and Applied Computer Science, ul. Orla 171, 30-244 Kraków, Poland}
\email{Agnieszka.Pollo@ncbj.gov.pl}

\author{L. Wang}
\affiliation{SRON Netherlands Institute for Space Research, Landleven 12, 9747 AD Groningen, The Netherlands}
\email{l.wang@sron.nl}

\author[orcid=0000-0002-9495-0079]{V. Rodriguez-Gomez}
\affiliation{Instituto de Radioastronom\'ia y Astrof\'isica, Universidad Nacional Aut\'onoma de M\'exico, Apdo. Postal 72-3, 58089 Morelia, Mexico}
\email{vrodgom.astro@gmail.com}

\author[orcid=0000-0002-7217-5120]{A. La Marca}
\affiliation{SRON Netherlands Institute for Space Research, Landleven 12, 9747 AD Groningen, The Netherlands}
\affiliation{Kapteyn Astronomical Institute, University of Groningen, Postbus 800, 9700 AV Groningen, The Netherlands
}
\email{A.La.Marca@sron.nl}

\begin{abstract}
Galaxy mergers are crucial for understanding galaxy evolution, and with large upcoming datasets, automated methods, such as Convolutional Neural Networks (CNNs), are essential for efficient detection. It is understood that CNNs classify mergers by identifying deviations from the regular, expected shapes of galaxies, particularly faint features, which are indicative of a merger event. In this work, we present a novel investigation of the relative importance of different morphological components: faint residual features and position and spatial structure, in CNN-based binary classification of galaxies into merger and non‑merger classes. Using mock images from the IllustrisTNG simulations processed to mimic Hyper Suprime-Cam (HSC) observations, we fit S{\'e}rsic profiles to each galaxy and generate three datasets: original images, model images containing only smooth S{\'e}rsic profiles, and residual images highlighting faint features after model subtraction. We train three identical CNNs on these datasets: CNN1 on original images, CNN2 on model images, and CNN3 on residual images. CNN1, trained on full images, achieves the highest accuracy of 74\%. CNN2, using only shape information including source position, achieves 70\%, while CNN3, using only faint residual features, achieves 68\%. We find that galaxy merger classification is possible using either faint features or the position and S{\'e}rsic profile information present in residual and model images, respectively. Our results demonstrate that not only faint features but also source position information play complementary roles in merger classification. This has important implications for the design and interpretation of machine learning methods for galaxy morphology, particularly in regimes where specific image components may be enhanced or suppressed.


\end{abstract}

\keywords{Galaxies: mergers --- Galaxies: faint structures --- Galaxies: classification --- Methods: machine learning --- Techniques: image processing --- Techniques: S{\'e}rsic profile fitting}

\section{Introduction}
In the Lambda-cold dark matter cosmology, dark matter (DM) halos form due to gravitational instability, allowing baryonic matter to cool, accumulate, and form galaxies. These galaxies then grow through both accretion within these halos and mergers. Hence galaxy mergers provide crucial insights into how galaxies evolve over time. The merging process proceeds differently depending on the properties of colliding galaxies. Major mergers, where the stellar mass ratio of the merging components is up to 4:1, can combine their structures, resulting in a significant change in morphology if the components collide at the right speed and angle \citep{bournaud2005galaxy,lotz2010effect}. Conspicuous merger features, such as bridges, double nuclei, and tidal features (tails and plumes) of different sizes and shapes may appear \citep{toomre1972galactic}.
This process can feed supermassive black holes, activating galactic nucleus (AGN) activities \citep{gao2020mergers,di2005energy}, and trigger star formation \citep{pearson2019effect,ellison2013galaxy}. When the mass of one of the merging galaxies is significantly larger than the mass of the other one (minor mergers), the merging process is less violent, and the less massive galaxy is usually incorporated into the larger one, leaving the larger galaxy fairly undisturbed.

The galaxy merger fraction has evolved significantly during cosmic time. It is believed that mergers account for less than 10\% of the low-redshift galaxies \citep{ferreira2020galaxy,mundy2017consistent,duncan2019observational}, with the percentage increasing to 20\% within $z\in[2,3]$ \citep{tasca2014evidence}. The merger fraction decreases to 5\% at $z\approx 6$ \citep{ventou2019new}. Merger rates in Illustris \citep{vogelsberger2014introducing} simulations align with observational rates for medium-sized galaxies \citep{rodriguez2015merger}.

It is possible to use visual inspection \citep{darg2010galaxy} to identify a merger, though this method is challenging to replicate consistently, time-consuming for large samples of galaxies, and may suffer from low accuracy, incompleteness \citep{Huertas-Company}, and human biases \citep{lambrides2021merger}. Using morphological parameters is a way to quantify morphology with reliable and reproducible measurements. Concentration \citep{okamura1984digital,kent1985ccd,conselice2003relationship}, asymmetry \citep{conselice2000asymmetry}, and smoothness \citep{takamiya1999galaxy,conselice2003relationship}  (CAS parameters) and the bulge-merger statistics - a combination of the Gini coefficient \citep{abraham2003new} and $M_{20}$ (the second-order moment of the brightest 20\% of the galaxy) \citep{lotz2004new} - serve as feature detectors and are often used for merger classification \citep{okamura1984digital,abraham1996morphologies,bershady2000structural,snyder2015diverse}. Deriving those parameters requires images with sufficiently high spatial resolution. \cite{conselice2000asymmetry} has shown that the accuracy of CAS parameters begins to decline below the threshold value of  $1 \,\textrm{kpc}\, \textrm{pixel}^{-1}$. \cite{pawlik2016shape} introduced the shape asymmetry parameter, a morphological indicator specifically designed for automated detection of galaxies in post-coalescence merger stages by quantifying asymmetric tidal features in galaxy outskirts. Another commonly used method to identify mergers is the close pair method, which selects galaxy pairs that are close on the sky and in redshift \citep{barton2000tidally,alonso2004galaxy,de2005millennium,duncan2019observational}. This method can use both spectroscopic and photometric observations\citep{lin2008redshift}. While spectroscopic methods provide greater precision than photometric techniques \citep{hildebrandt2008blind}, they are also more resource-intensive.
 Identifying galaxy mergers may also be possible using  only photometric information. \cite{suelves2023merger} found that photometric bands, colours, and their errors provide accurate classification of SDSS observations. 


With the advent of upcoming surveys such as \textit{Euclid} \citep{laureijs2011euclid} and  Legacy Survey of Space and Time \citep[LSST;][]{ivezic2019lsst} that will collect large amount of data it becomes even more important to create algorithms to automate time-consuming, repetitive tasks such as identifying galaxy mergers. The application of machine learning (ML) in the field of astronomy has experienced a significant surge over the past few years. Recent studies have shown that it is possible to use Convolutional Neural Networks (CNN) to address the visual merger classification problem.        \cite{ackermann2018using} used transfer learning to fine-tune CNN trained to classify real life objects, using SDDS observations.  \cite{pearson2019identifying} trained one CNN with observations from SDSS and another with simulated galaxies from EAGLE observations receiving an accuracy of 91.5\% and 65.2\% respectively. Combination of CNNs and Linear Networks has led to creation of the North Ecliptic Pole merging galaxy catalogue \citep{pearson2022north} for $z<0.3$. 
 \cite{ciprijanovic2020deepmerge} used a CNN to distinguish between mergers
and non-mergers in simulated images from Illustris-1 at $z=2$ with 79\% accuracy. \cite{bickley2021convolutional} showed that CNNs outperform Gini-M20 and asymmetry methods in post-merger and isolated galaxy classification.

The performance of these networks vary due to many factors, including differences in setup. \cite{margalef2024galaxy} conducted a systematic and consistent study between machine learning-based detection methods to understand the relative performance of different ML methods using the same training and test data. A total of six ML methods have been tested, based on the same mock IllustrisTNG simulation images generated to mimic the HSC data. All CNNs that were not pre-trained with other galaxy images show similar results, using different architectures. This may indicate that preprocessing the training data is a more important factor compared to the choice of parameters of the CNNs. The above works do not exhaust the issue of classifying mergers with CNNs and there are also other works on this topic. 
\citep[e.g.][]{walmsley2019identification,wang2020towards,ferreira2020galaxy,nevin2019accurate,sanchez2023identification}

The previous work on creating a CNN classifier for classifying merger galaxies relied on image preprocessing limited to augmentation and normalization \citep{ackermann2018using,pearson2019identifying,pearson2022north,ciprijanovic2020deepmerge,bickley2021convolutional}. The wide range of characteristic features of mergers and their irregularity, especially subtle peculiarities, may cause these features to be inadequately captured by the network or get mixed up with other galaxy structures. \citet{mantha2019studying} explored the use of  S{\'e}rsic profile subtraction to extract merging features. They developed a tool that extracts and quantifies residual substructures hosted by galaxies. This tool can help to clarify how merger signatures are identified, could enhance efforts to connect observed galaxy morphology to the underlying physical processes of merging, and improve merger detection techniques. In particular, a fully automatic merger‑finding method based on the morphology of residual images after Sérsic model subtraction demonstrates that residual‑based diagnostics can effectively isolate disturbed systems in large samples \citep{hoyos2012new}. 

In this work, we create model images for each galaxy using S{\'e}rsic profile fits applied to every image. We then subtract these model images from the originals to produce the residual images. This process leaves behind everything that does not match the profile, possibly emphasizing faint merger peculiarities such as diffuse structures or tidal features. It is believed that they are of great importance for the classification \citep{pearson2022north}, so we want to explore whether CNNs can be trained on these faint structures. Rather than simply attempting to improve overall classification accuracy, our approach aims to provide physical insight into what information CNNs utilize for merger identification. By isolating different morphological components through S{\'e}rsic profile fitting and subtraction, we can quantify the relative importance of bulk galaxy properties versus faint tidal features. This work tests and compares CNNs trained on three distinct datasets:
\begin{itemize}
    \item CNN 1, trained on original images,
    \item CNN 2, trained on S{\'e}rsic profile images,
    \item CNN 3, trained on residual images.
\end{itemize}

This paper is structured as follows. In Sect. \ref{data} we present the IllustrisTNG simulations data. In Sect. \ref{methodology} we describe the data preprocessing, the models' architecture, and define performance metrics used for evaluation. In Sect. \ref{results} we present performance of the models and study the models behaviour. In Sect. \ref{discusion} we compare the performance of the models in merger classification task. In Sect. \ref{conclusions}, we present our conclusions and future work. We assume a cosmology consistent with the Planck \cite{Planck2016} results
($\Omega_{\Lambda,0} = 0.6911$, $\Omega_{m,0}= 0.3089$, $h = 0.677$).


\section{Data} \label{data}
\subsection{Illustris TNG}
Images used for this study
come from Next Generation Illustris Simulations  \citep[IllustrisTNG, or TNG;][]{nelson2019illustristng}, a cosmological gravo-magnetohydrodynamical simulations of formation and evolution of galaxies assuming cosmological parameters consistent with Planck \cite{Planck2016}. TNG comprises of three runs, TN50 TNG100, and TNG300, with comoving box
sizes of 50, 100, and 300 Mpc $h^{-1}$, respectively. All runs follow the same framework. Star formation follows the  Springel \& Hernquist model \cite{springel2003cosmological}. Stars are formed stochastically  after gas reaches the density threshold of $n=0.1H cm^{-3}$. The formation process obeys the Kennicutt-Schmidt relation \citep{kennicutt1998global} and \cite{chabrier2003galactic} initial mass function. Stars evolve and feed back the mass to the interstellar medium via Type Ia and Type II supernovae. Presence of a redshift-dependent, spatially uniform, ionizing UV background radiation field cools metal-enriched gas. The radiation field of nearby AGN activities further adjust the cooling process by superimposing it with the UV background. When a halo is sufficiently massive, a super massive black hole (SMBH) forms, accretes gas from the nearby area and contributes energy back into its surroundings. For a more comprehensive understanding of IllustrisTNG, we direct the reader to the work of \cite{Pillepich_2017}.

 The merger trees were constructed by directly tracking the baryonic content of the subhalos \cite{rodriguez2015merger}. They were later used to create a dataset of merging galaxies by \citet{margalef2024galaxy}. They selected galaxies from TNG100 and TNG300. The dataset consists of galaxies in the redshift range $[0.1, 1.0)$. Merging galaxies either had a merger event in the past 0.3 Gyr or will have one in the next 0.8 Gyr. \cite{margalef2024galaxy} only consider major mergers where the stellar mass ratios are greater than 4:1. Additionally, they categorize mergers that are -0.8 to -0.1 Gyr before coalescence as pre-mergers, -0.1 Gyr before to 0.1 Gyr after as on-going mergers, and 0.1 to 0.3 Gyr after the coalescence as post-mergers.
\subsection{Mock images}

\begin{figure}[h]
    \centering
        \centering
        \includegraphics[width=\linewidth]{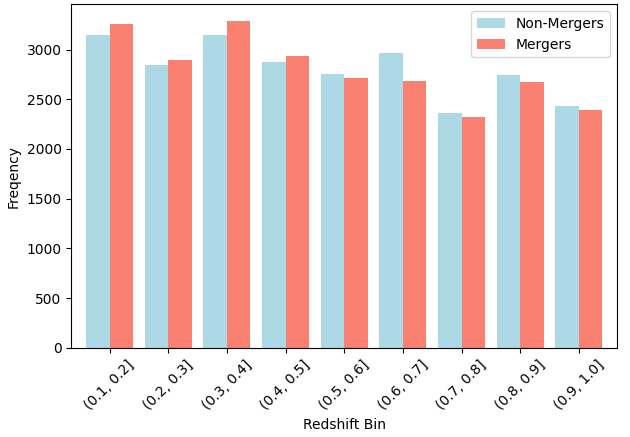}
        \caption{Distribution of galaxies across redshifts. 
        Y axis shows number of galaxies in each redshift bin and 
        X axis shows the redshift bin.}
        \label{dist_red}
    \hfill
        \centering
        \includegraphics[width=\linewidth]{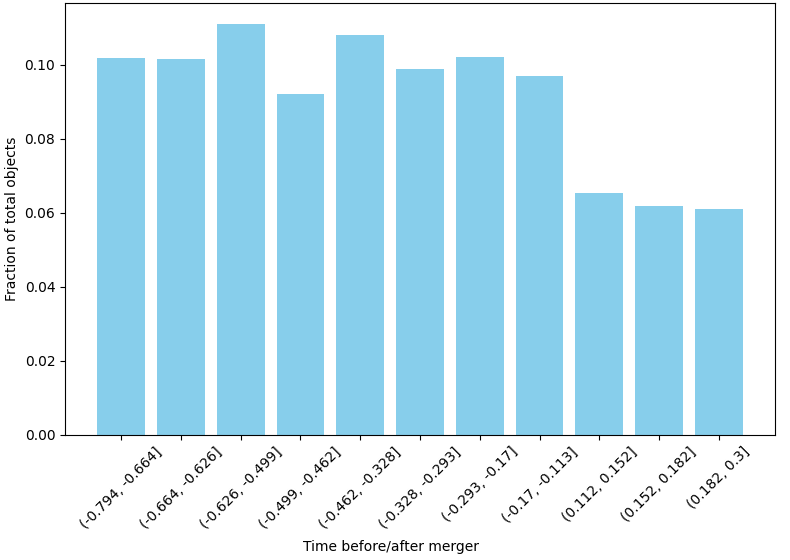}
        \caption{Distribution of galaxies across pre-merging and post-merging times. 
        Y axis shows fraction of galaxies and 
        X axis shows merging time bins in Gyr.}
        \label{dist_time}
\end{figure}

Simulated galaxy images are then processed to mimic HSC observations as described in \cite{margalef2024galaxy}. The sum of spectral energy distribution of all stellar particles in the image is passed through the filter to create a projected 2D map. Then each image is convolved with HSC i-band PSF. On top of that, Poisson noise that occurs in optical devices is added, and the image is injected into cutouts of real HSC observations. For a more complete description of how the synthetic images were generated, we direct the reader to \citep{rodriguez2019optical}.
A physical size of the images was set to 160 kpc as it is approximately the typical separation between binary mergers based on simulations \citep{qu2017chronicle,moreno2019interacting}. Images are 320x320, 192x192, 160x160, and 128x128 pixels for $[0.1, 0.31)$, $[0.31, 0.52)$, $[0.52, 0.76)$, and $[0.76, 1.0)$ redshift bins respectively. We randomly divided the images into three sets in the 75:15:10 ratio. The training set, used to fit the parameters, the validation set, providing an evaluation of a model while tuning the model’s hyperparameters, and finally the test set, used for an unbiased evaluation of the final model. Galaxies were divided, so that galaxies belonging to the same merger sequence are included in only one of the three sets.
The distribution of galaxies across redshifts can be seen in Fig. \ref{dist_red}, and the distribution of galaxies across merging times in Fig. \ref{dist_time}. The final mock sample contains $502\,152$ galaxies, of which 
$249\,679$ are labelled as mergers and 
$252\,473$ as non-mergers. Among the mergers, 
$174\,689$ are pre-mergers, 
$34\,574$ are ongoing mergers, and 
$40\,416$ are post-mergers.
The distribution of systems over these classes is summarized in Table~\ref{numery}.

\begin{table}[h!]
\centering
\begin{tabular}{l r}
\hline
Mergers and non-mergers       & 502\,152 \\
Mergers  & 249\,679 \\
Non-mergers  & 252\,473 \\
Pre-mergers & 174\,689\\
On-going mergers & 34\,574\\
Post-mergers & 40\,416 \\

\hline
\end{tabular}
\caption{Summary of the galaxy counts in the full sample, including mergers, non-mergers, and the subdivision of mergers into pre-mergers, ongoing mergers, and post-mergers.}
\label{numery}
\end{table}

\section{Methodology} \label{methodology}

\subsection{S{\'e}rsic profile fitting}
We used statmorph \citep{rodriguez2019optical} - an open source Python package for calculating
 morphologies of galaxy images to fit 2D S{\'e}rsic profile of each galaxy in the image. First, we create a segmentation map using the SEP Python package \citep{barbary2016sep} - code derived from the Source Extractor \citep{sex} code base.
A segmentation map is a map the same size as the original image in which every pixel has an assigned label such that pixels with the same label belong to the same source, and defines the area over which morphological parameters are calculated. Since each pixel value is a combination of background and light from objects, SEP estimates the background level at every point in the image and subtracts it from the original image to accurately measure the flux of the objects. With background-subtracted data SEP performs object detection. We set the detection threshold to a value of $1.5\sigma$ where $\sigma$ is the global background root mean square and minimum number of pixels required for an object is set to 5. Then overlapping but different sources are separated using a deblending algorithm to isolate the galaxy of interest for calculations.



For each galaxy in the image we perform photometric measurements and estimate the S{\'e}rsic parameters (half light radius, surface brightness at half-light radius, S{\'e}rsic index, position of the center of the galaxy, ellipticity of the profile, and rotation angle). The estimated S{\'e}rsic parameters serve as an initial guess for the S{\'e}rsic parameters used to create model images of the galaxies. Providing good initial guesses is crucial because the Levenberg–Marquardt algorithm \citep{Geda_2022} used for fitting is sensitive to initial parameters and, if provided with bad initial guesses, it may return parameters that correspond to a local minimum.

\begin{figure*}[htbp]
    \centering
    \includegraphics[width=\textwidth, height=0.4\textheight, keepaspectratio]{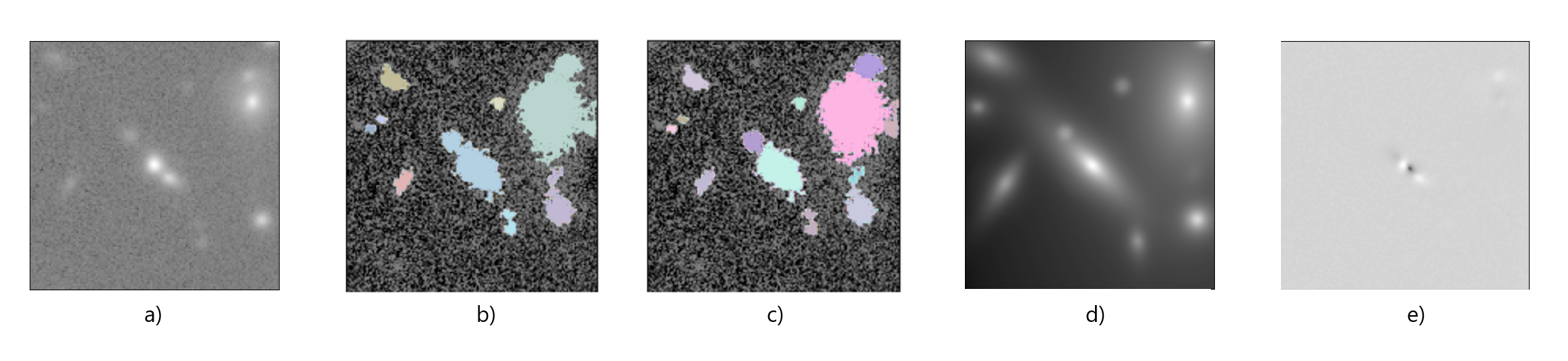}
    \caption{Preprocessing steps. a) Original image, b) segmentation map, c) deblending to isolate overlapping galaxies, d) S{\'e}rsic profile fits, and e) subtraction to produce residual image.}
    \label{steps}
\end{figure*}

\begin{figure}[htbp]
    \centering
    \begin{minipage}{0.94\columnwidth}
        \centering
        {\small
        \begin{tabular*}{\columnwidth}{@{\extracolsep{\fill}}ccc}
        Original images & Model images & Residual images \\
        \end{tabular*}
        }
        \vspace{0.3em}
        \includegraphics[width=0.94\columnwidth, keepaspectratio]{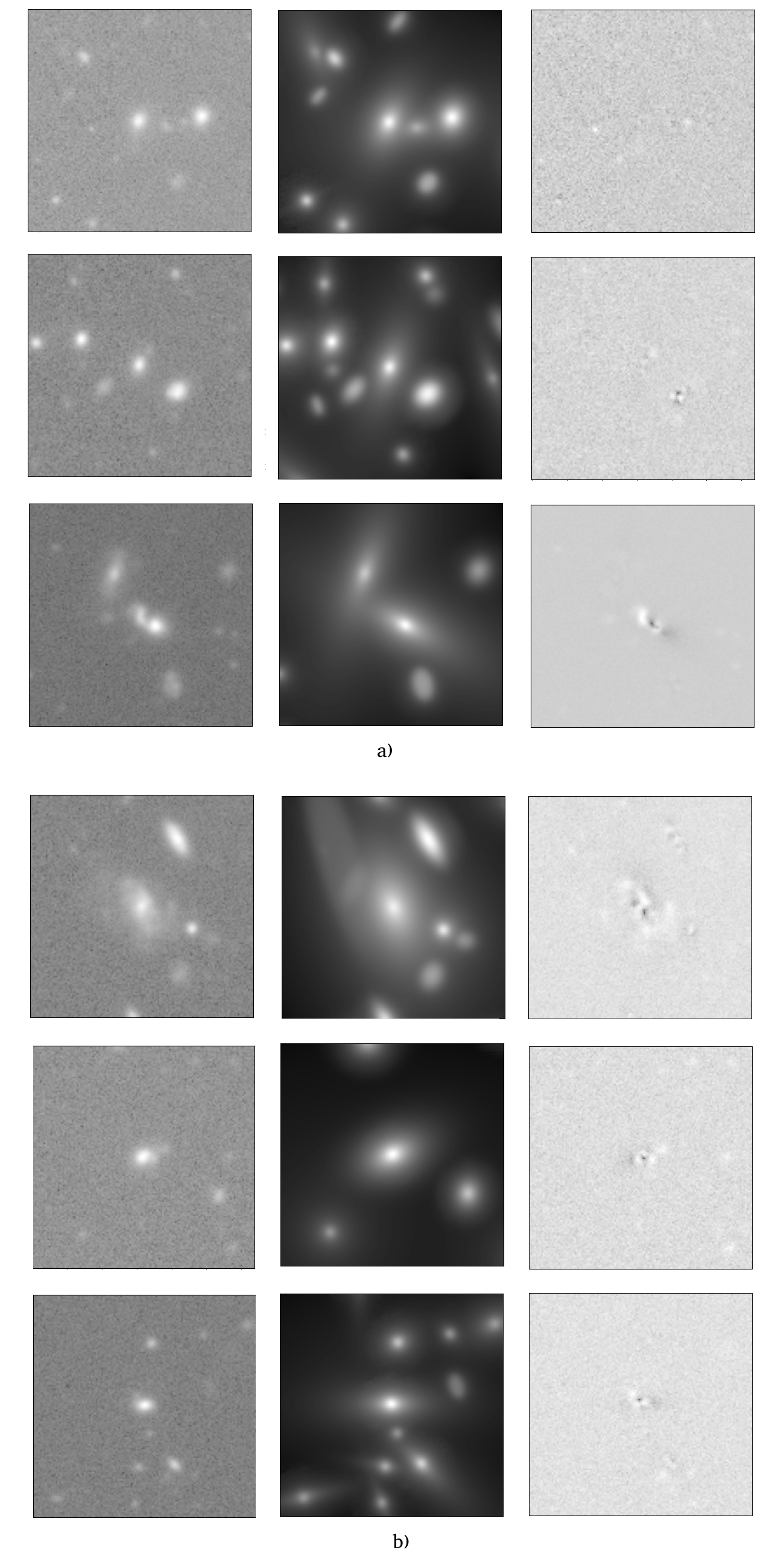}
    \end{minipage}
    \caption{Examples of mergers (a) and non-mergers (b) used for training. Left: original images. Middle: fitted S{\'e}rsic profiles. Right: residual images. A power-law normalization with exponent 0.2 was used to display the images.}
    \label{NonmerTrain}
\end{figure}

S{\'e}rsic profiles of each galaxy in every image were subtracted from the original image, creating a residual image, and cropped to 128 x 128 pixels. Resizing the image is done after subtraction of S{\'e}rsic profiles from original sized image to ensure the good quality of the residuals on the edges of the images. The steps used to create training datasets are shown in Fig. \ref{steps}. Examples of non-merging and merging galaxies in the dataset are shown in Fig. \ref{NonmerTrain}.

\subsection{S{\'e}rsic profile fitting versus masking approaches for feature isolation}

While masking approaches, such as removing central pixels or high-flux regions, can be used to isolate faint features in galaxy images, S{\'e}rsic profile fitting offers several important advantages for the analysis performed in this work. By modeling and subtracting the smooth, symmetric light distribution of the galaxy with a S{\'e}rsic profile, this method preserves the full radial structure of the galaxy and avoids the artificial truncation and edge artifacts that masking techniques often introduce. Masking can inadvertently remove or distort low-surface-brightness features, which are critical for identifying tidal structures and other merger signatures, and may also eliminate double nuclei or other key features relevant for pre-merger identification.

\subsection{Deep learning}
\subsubsection{Convolutional neural networks}
In this work we use a standard CNN for image‑based binary classification of mergers and non‑mergers. CNNs combine convolutional, pooling, and fully connected layers with non‑linear activations and regularization techniques such as batch normalization and dropout \citep{nair2010rectified,lecun2002gradient,ioffe2015batch,wu2015towards}; a full, detailed description of machine learning is beyond the scope of this paper.

\subsubsection{Training and architecture}
\begin{figure}[h]
   \centering
   \includegraphics[width=\columnwidth]{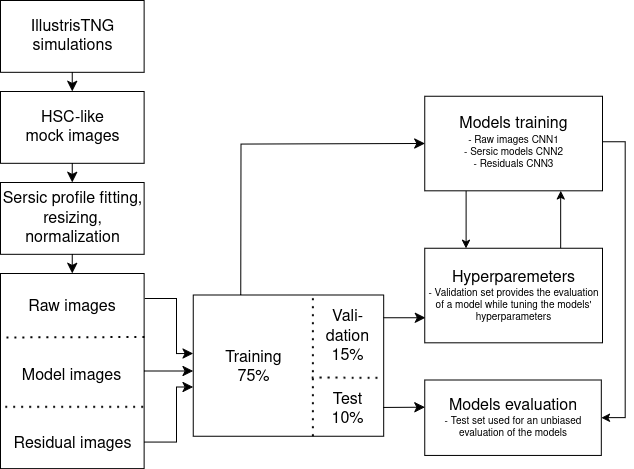}
   \caption{Data pipeline summary. Three datasets were created  -  raw images, S{\'e}rsic profile images, and residual images. Images were cropped to 128 × 128 pixels, linearly normalized, and divided in a 75:15:10 ratio into the training set, the validation set, and the test set.}
   \label{pipeline}
\end{figure}

The raw images, model images, and residual images are linearly normalized in a range of 0 to 1. These preprocessed images serve as input for all three CNNs. The architecture is adapted from the 'Method‑5 (CNN2)' configuration of \cite{margalef2024galaxy}, with a small change in kernel size. The output of the network is a single neuron activated by a sigmoid function that provides values representing the likelihood that the image contains a merger. 20\% dropout was used before max-pooling layers \citep{wu2015towards}.
The loss of the network was determined using the binary cross-entropy function and was optimized with the ADAM \citep{diederik2014adam} algorithm with learning rate $\alpha = 3 \cdot 10^{-4}$. The CNN has 8\,966\,273 total trainable parameters. All networks were trained for 50 epochs. All three networks showed signs of overfitting after around 15 epochs, where training accuracy kept increasing while validation accuracy remained relatively still. It means that the problem does not take longer then 50 epochs to train. We used the set of tuned weights and biases for the epoch that provided the highest value of validation accuracy before showing signs of overfitting for classification for the threshold for which precision is equal to recall. A summary of the data pipeline from the creation of the data to the training of the model is shown in Fig. \ref{pipeline}, and the summary of CNN architecture in Table \ref{architecture_table}. The CNNs were implemented using the PyTorch \citep{paszke2017automatic} deep‑learning library in Python. Training was performed on RTX 4090 GPU. Training a single CNN for 50 epochs on the full training set required approximately 2 hours, and the total training time for all three models was approximately 6 hours.

\subsubsection{Evaluation metrics}
In binary classification problems, a model can correctly or incorrectly assign a class for a given input. True
positive (TP) is an object of positive label (merger) that is correctly classified, and true negative (TN) is an object of
negative label (non-merger) that is correctly classified. Similarly, a false positive (FP) is an image of the actual negative label classified
as positive and false negative (FN) and vice versa. 
To evaluate an already trained model, we use several metrics such as accuracy, precision, F1 score, and receiver operating characteristic curve (ROC-curve). Accuracy is the number of correct predictions divided by the total number of predictions:
\begin{equation*}
    \text{Accuracy}=\frac{TP+TN}{TP+TN+FP+FN}\quad.
\end{equation*}
Precision is the number of correctly identified positive cases divided by the total number of predicted positive cases,
\begin{equation*}
    \text{Precision}=\frac{TP}{TP+FP}\quad.
\end{equation*}
Recall is equal to the number of correctly predicted positive cases divided by the total number of true positive cases,
\begin{equation*}
    \text{Recall}=\frac{TP}{TP+FN}\quad.
\end{equation*}
The F-1 score is the harmonic mean of the precision and the Recall:
\begin{equation*}
    \text{F-1}=\frac{2}{Precision^{-1}+Recall^{-1}}\quad.
\end{equation*}
The last metric used is the area under the curve (AUC). The receiver operating characteristic curve (ROC) shows True Positive Rate (TPR) against False Positive Rate (FPR) at different decision thresholds. TPR is the proportion of actual positive cases correctly identified by the model, while the FPR is the proportion of actual negative cases incorrectly identified as positive. AUC is the area under
this curve and ranges in value from zero to one. A model whose predictions are always wrong has an AUC of 0 and a model whose predictions are always correct has an AUC of 1. For a random classifier, AUC = 0.5, ROC is a diagonal line and a model has no capacity to distinguish between positive and negative class.

\section{Results} \label{results}
\begin{figure}[tbp]
    \centering
        \centering
        \includegraphics[width=\linewidth]{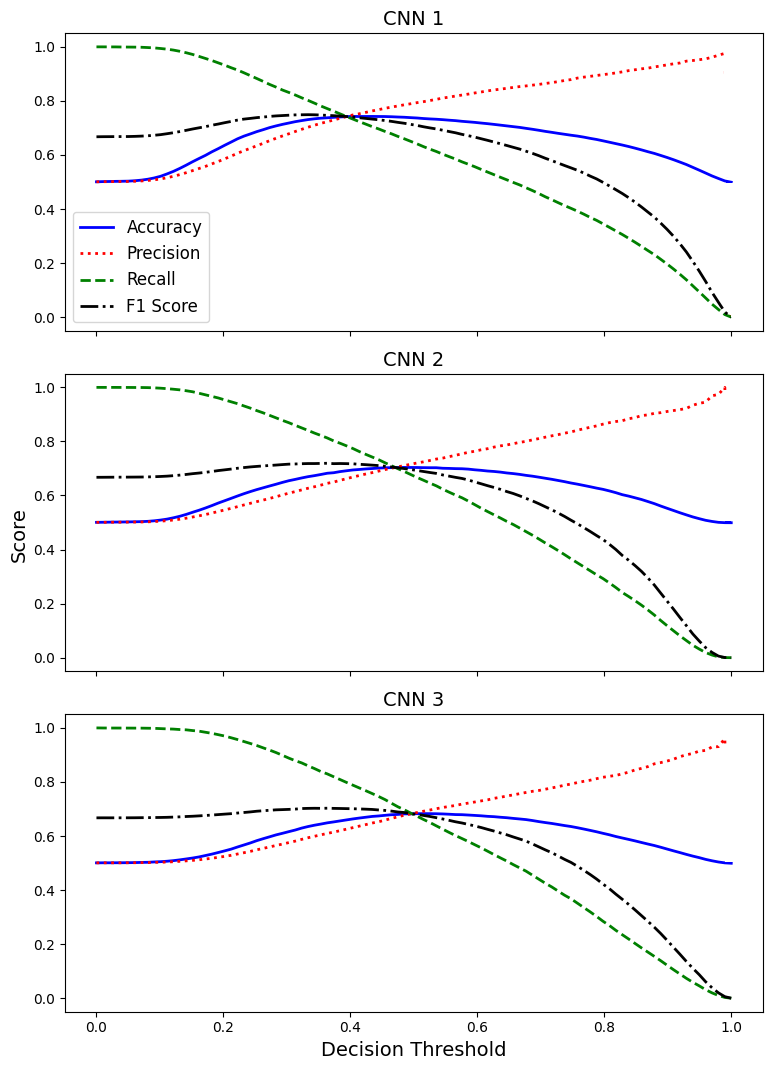}
        \caption{Performance metrics as a function of threshold of CNN1 (top panel), CNN2 (middle panel), and CNN3 (lower panel). Accuracy is shown with blue solid line. Precision with red dotted line. Recall with green dashed line, and F1 with black dashed dotted line.}
        \label{raw_metrics}
\end{figure}

\begin{figure}[tbp]
    \centering
        
        \includegraphics[width=\linewidth]{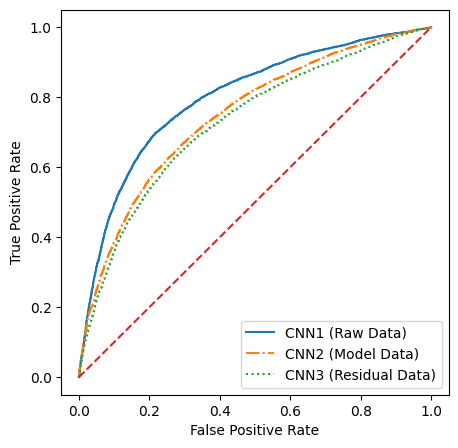}
        \caption{ROC curves of all networks. This curve plots TPR against FPR at different classification thresholds. CNN1 is shown with blue solid line, CNN2 with yellow-green dotted line, and CNN3 with dotted line. The ROC curve of a random classifier is shown with a dashed red diagonal.}
        \label{roc_all}
        \centering
        \includegraphics[width=\linewidth]{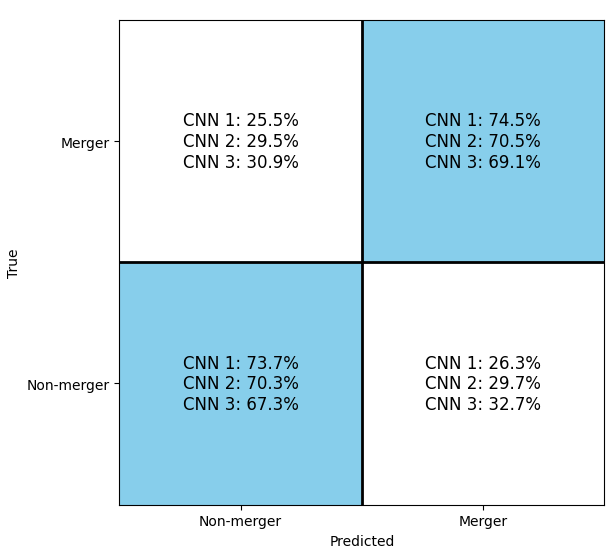}  
        \caption{Confusion matrix of all three CNNs normalized by row. This shows the proportion of predictions for each class.}
        \label{raw_cf}
\end{figure}

We show the performance of all three networks using the evaluation metrics described in Sect. \ref{methodology}. Evaluation metrics scores depend on the choice of a decision threshold as it changes the required score to classify galaxy as a merger. Here, we set the threshold for each CNN independently, but the change in metrics as a function of threshold was also examined in Fig. \ref{raw_metrics}. For all CNNs, we choose the decision threshold at which precision equals recall to achieve a balanced trade-off between the rate of correctly identified positives and the rate of false positives, as both false positives and false negatives are equally undesirable. For CNN1 (raw images), CNN2 (model images), and CNN3 (residual images) it is 0.39, 0.47, and 0.49 respectively. CNN1 and CNN2 achieve their peak accuracy for these thresholds. The decline in accuracy after the peak is slower for CNN3 than CNN1 and CNN2. The overall performance across the entire redshift range for all networks is summarized in Table \ref{metrics_all}. The results show that the CNN1 performs the best in terms of accuracy -  74\%. CNN2 and CNN3 have similar overall accuracy - correctly classifying about, respectively, 70\% and 68\% of all images. ROCs of all CNNs are shown in Fig. \ref{roc_all}, and confusion matrices in Fig. \ref{raw_cf}. The overall best network is CNN1, followed by CNN2, and CNN3. CNN1 performs better in recognizing merging galaxies. 74.5\% of them are correctly identified, while for non-merging galaxies, this number drops to 73.7\%. CNN2 and CNN3 show similar performance. CNN2 correctly predicts 70.5\% of mergers and 70.3\% of non-mergers, while CNN3 correctly predicts 69.1\% of mergers and 67.3\% of non-mergers. 
\begin{table}[h]
\centering
\caption{Overall performance of CNN1 (raw images network), CNN2 (S{\'e}rsic profiles network), and CNN3 (residual images network).}
\label{metrics_all}
\vspace{0.5em}
\begin{tabular}{c|c|c|c|c}
\hline
Method & Accuracy & Precision & Recall & F1 \\ \hline
CNN1   & 74\%     & 73\%      & 74\%   & 74\% \\
CNN2   & 70\%     & 70\%      & 70\%   & 70\% \\
CNN3   & 68\%     & 68\%      & 69\%   & 68\% \\ \hline
\end{tabular}
\end{table}



\subsection{Redshift}

\begin{figure}[htbp]
        \centering
        \includegraphics[width=\linewidth]{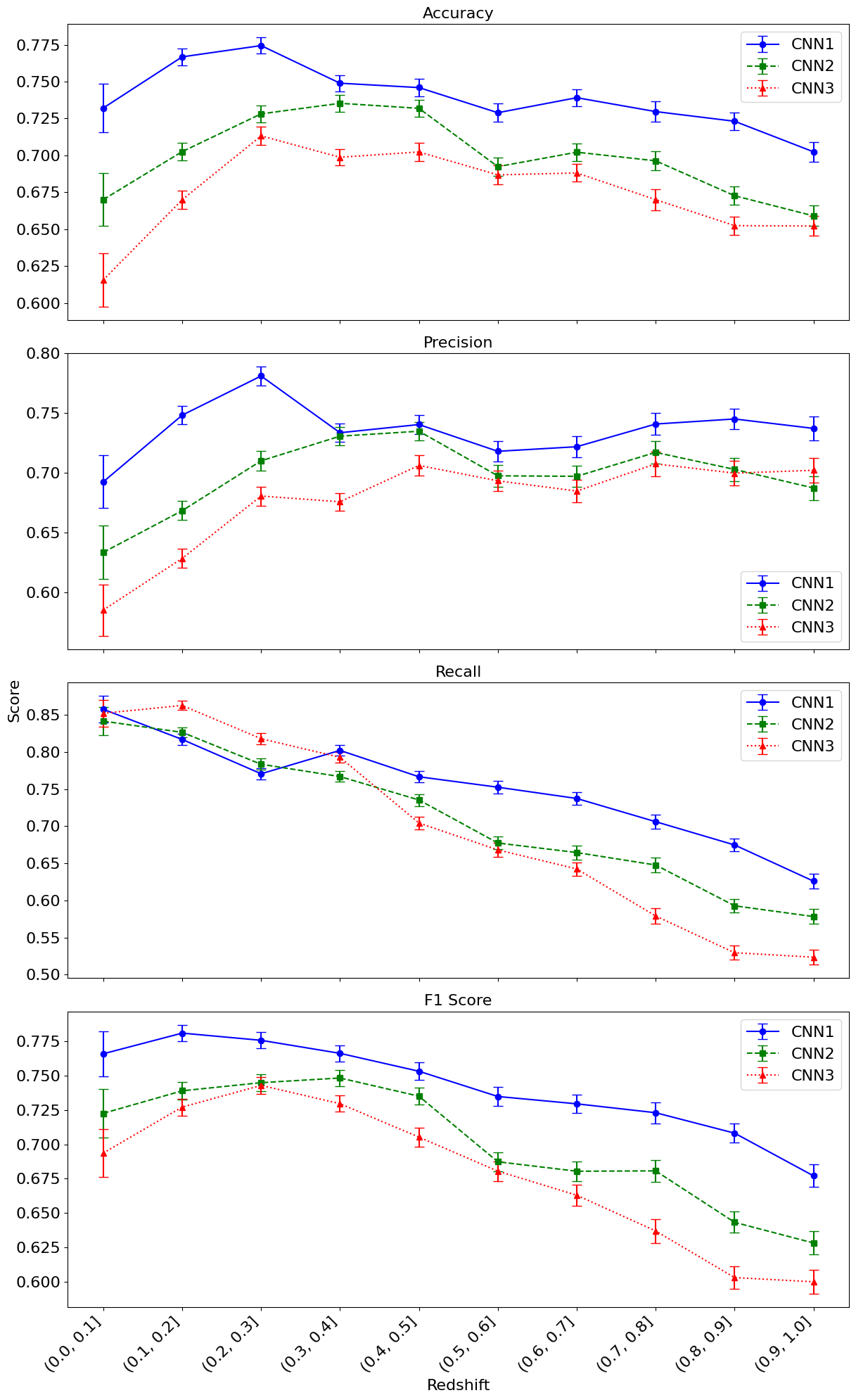}
        \caption{Metric values across redshifts with standard deviation. From top to bottom: accuracy, precision, recall, and F1. Metrics for CNN1 are shown with blue solid line and circles, for CNN2 with green dashed line and squares, for CNN3 with red dotted line and triangle facing upwards.}
        \label{raw_metrics_red}
\end{figure}

\begin{figure}[tbp]
    \centering
    \includegraphics[width=\linewidth]{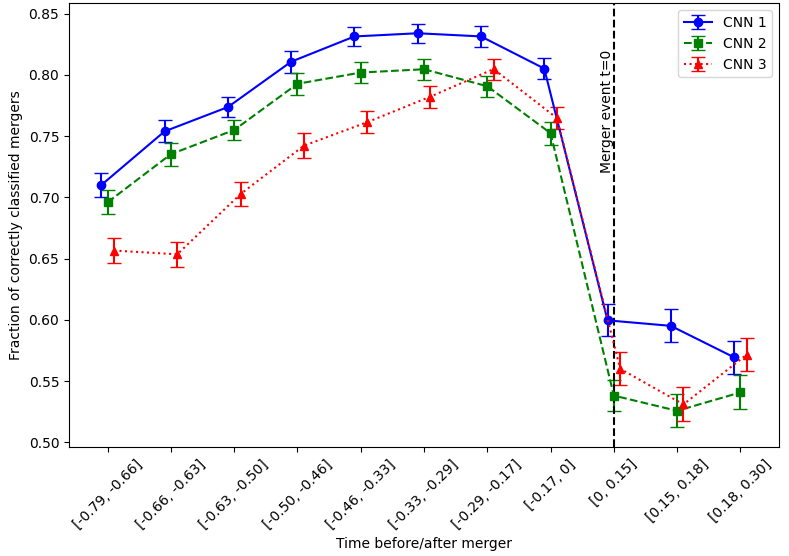}
    \caption{Fractions of correctly classified mergers across merger times in Gyr with standard deviation. CNN1 is shown with blue solid line and circles. CNN2 with green dashed line and squares. CNN3 with red dotted line and triangles facing upwards.}
    \label{preds_bins}
\end{figure}

We check how the performance of all models changes as a function of redshift in Fig. \ref{raw_metrics_red}, and across merging time bins in Fig. \ref{preds_bins}. We performed bootstraping \citep{bootstraping}, a resampling technique, in which multiple new samples (in our case 1000 resamples for each bin) are drawn with replacement from the original data to estimate uncertainties.


All three networks show similar behavior across the redshift bins. As expected, the accuracy gradually decreases with redshift due to the deterioration of spatial resolution at higher redshifts, with the exception of the first redshift bins, where we observe an increase in accuracy for redshifts 0.1 to 0.3 for CNN1 and CNN3, and for redshifts 0.1 to 0.4 for CNN2. This is most likely because these images were originally larger (320x320 pixels) and resized to 128x128, potentially causing some loss of information. 
\subsection{Merger time}
As seen by the higher fraction of correctly identified mergers, CNN1 achieves the best performance across most of the examined time intervals (-0.79 Gyr to 0.18 Gyr). CNN3 classifies mergers with similar precision as CNN1 in the later phase of the merging process (0.18 to 0.3 Gyr). As summarized in Table \ref{corrprepost}, CNN1 correctly classifies $79.3 \pm 0.6\%$ of pre-mergers and $58.8 \pm 1.5\%$ of post-mergers, CNN2 correctly classifies $76.6 \pm 0.6\%$ of pre-mergers and $53.4 \pm 1.6\%$ of post-mergers, and CNN3 correctly classifies $73.3 \pm 0.6\%$ of pre-mergers and $55.5 \pm 1.4\%$ of post-mergers. Reported values are the mean fractions of correctly classified post-mergers and pre-mergers for each model, with the error represented as the difference between the mean and the upper bound of the 95\% confidence interval of 1000 bootstrap resamples. The substantially lower fractions of correctly classified post‑mergers compared to pre‑mergers for all three CNNs are likely influenced by the relative scarcity of post‑mergers in the training data. As seen in Table \ref{numery}, the underlying mock sample contains fewer post‑mergers than pre‑mergers, which tends to bias the networks toward the better‑represented pre‑merger morphologies. This imbalance, in combination with the intrinsically fainter features expected in post‑mergers, probably contributes to the systematically poorer performance of all models in this regime.

\begin{table}[h]
\centering
\caption{Fraction of correctly classified pre-mergers and post-mergers for all CNNs.}
\label{corrprepost}
\vspace{0.5em}
\begin{tabular}{c|c|c}
\hline
Method & Frac. pre-mergers & Frac. post-mergers \\ \hline
CNN1   & $79.3 \pm 0.6\%$  & $58.8 \pm 1.5\%$   \\
CNN2   & $76.6 \pm 0.6\%$  & $53.4 \pm 1.6\%$   \\
CNN3   & $73.3 \pm 0.6\%$  & $55.5 \pm 1.4\%$   \\ \hline
\end{tabular}
\end{table}
\subsection{Galaxy ellipticity}
\begin{figure}[tbp]
    \centering
    \includegraphics[width=\linewidth]{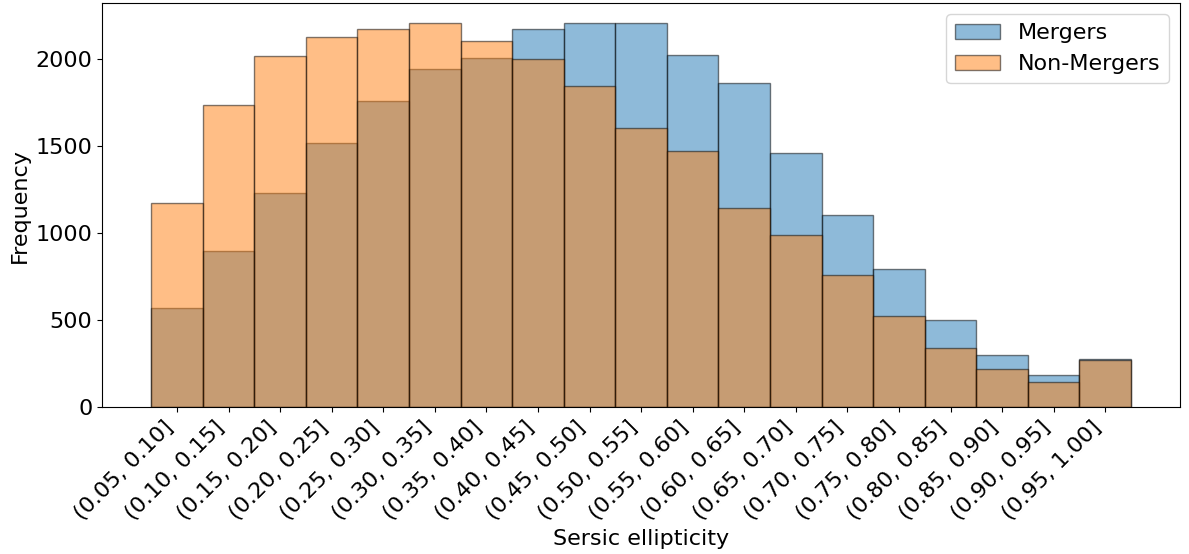}
    \caption{Distribution of mergers and non-mergers across ellipticities of the S{\'e}rsic profile for the central galaxy. Y-axis shows the number of galaxies and X-axis shows S{\'e}rsic ellipticity bins.}
    \label{sersicellip}
\end{figure}

As shown in Fig. \ref{sersicellip}, the distribution of S{\'e}rsic profile ellipticity for mergers and non-mergers is different. There are more non-mergers with low ellipticity than mergers. Statistically, mergers will have a more elliptical profile shape because they have undergone processes that could potentially distort their shape. Studying the dependence of metric values on the S{\'e}rsic profile shape will help us to understand to what extent the shape of a galaxy is useful in the evaluation of the network. 

\begin{figure}[htbp]
    \centering
        \centering
        \includegraphics[width=\linewidth]{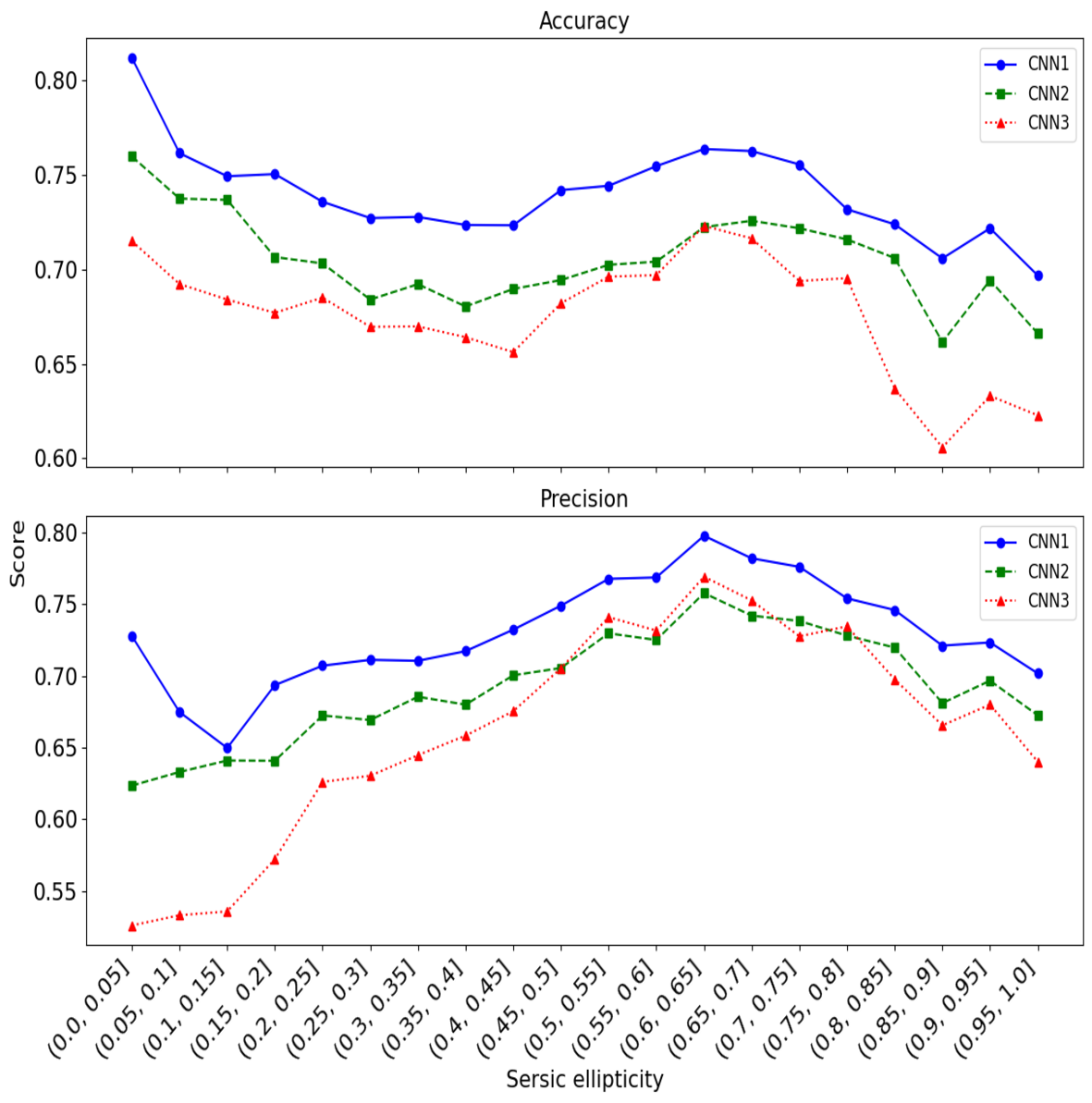}
        \caption{Accuracy (top) and Precision (bottom) across S{\'e}rsic ellipticities. Metrics for CNN1 are shown with a blue solid line and circles, for CNN2 with green dashed line and squares, for CNN3 with red dotted line and triangle facing upwards.}
        \label{metricsersicunbal}
        
\end{figure}

The metrics as a function of Sérsic ellipticity, shown in Fig. \ref{metricsersicunbal}, exhibit consistently high accuracy at low ellipticities for all three networks, followed by a general decline toward intermediate ellipticities and a mild recovery at (0.45, 0.65] bins. Precision rises more steadily: after an initial drop in the (0, 0.15] bins, it increases across most of the ellipticity range, reaching its peak around the (0.65, 0.70] bin before decreasing toward the highest ellipticities. These trends reflect the sensitivity of both metrics to the varying merger fraction within each bin - higher ellipticity bins contain relatively more mergers, which leads to more true positives and fewer false positives. These trends suggest that ellipticity provides useful discriminatory power for merger classification.

\section{Discussion} \label{discusion}
 \subsection{CNNs comparison}

 Among the three networks, CNN1 performs best overall, achieving the highest classification accuracy of 74\%. CNN1 has access to both shape and position information, as well as any residual features present in the data, which appears to be most effective for distinguishing mergers from non-mergers. CNN2, which uses only source position and shape information, performs slightly better (70\% accuracy) than CNN3 (68\% accuracy), which relies solely on residual features (see Table \ref{metrics_all} for overall performance of all CNNs). This suggests that source position and shape information play a somewhat more significant role in classification than residual features alone. The absence of such information, especially features like double nuclei that may be removed through Sérsic profile subtraction, can confuse the network and reduce its ability to accurately distinguish between mergers and non-mergers. 
 
 

 CNN1 was able to predict 79\% of pre-mergers correctly, a score similar to CNN2, which achieved 77\% accuracy in this task, and much higher than CNN3, which achieved 73\% accuracy, as seen in Table \ref{corrprepost}. This may suggest that source position and shape information is the most important factor in classifying pre-mergers. This would be expected because the CNNs are looking for galaxies close to the primary galaxy, and a lack of its presence strongly suggests there is no pre-merger event. However, it does not suggest that there is no post-merger event. We would expect the precision to increase gradually with pre-merging time, reaching its maximum around the merging event. This happens for CNN1 and CNN3 up to 0.17 Gyr before the merging event and up to 0.29 Gyr before the merging event for CNN2, as shown in Fig. \ref{preds_bins}.
 
Residual features are critical for classifying post-mergers. CNN2, which lacks residual information, has the lowest post-merger classification accuracy (53\%), while CNN3, which uses only residual information, performs slightly better at 56\%. CNN1, with access to all information, performs best with 59\% accuracy, as seen in Table \ref{corrprepost}. Over time, precision decreases for all networks as post-merger features fade. CNN3 consistently outperforms CNN2 across all post-merging times, suggesting that S{\'e}rsic profile subtraction helps detect fainter post-merger features. All three networks struggle more with distinguishing mergers from non-mergers in the post-merger stage than in pre-merger stages, possibly due to the faintness of residual features in post-merger galaxies. We can speculate that this is not due to the lack of double nuclei features or spatial information in general and also not due to the difference in the number of galaxies in each bin, as they are fairly similar for each pre- and post-merging stage as seen in Fig. \ref{dist_time}. A potential reason for this is that the residual features in the post-merger stage may be fainter and more difficult to detect.

The superior performance of CNN1 over CNN2 and CNN3 demonstrates that optimal merger classification requires integration of both morphological and residual information, with the modest performance gap between CNN2 (70\%) and CNN3 (68\%) confirming both components contribute substantially to classification accuracy. CNN1's consistent dominance across all redshifts (Fig.\ref{raw_metrics_red}), merger times (Fig. \ref{preds_bins}), and ellipticities (Fig. \ref{metricsersicunbal}) further indicates that morphological structure and residual features provide complementary information that neither component alone can fully replicate. However, both faint features and S{\'e}rsic profile information are useful for classification; otherwise, CNN2 and CNN3 would not perform significantly better than random classification.

The results in terms of accuracy are consistent with \cite{margalef2024galaxy}, who conducted similar studies on the same datasets using different methods, achieving accuracies of 70.5\%–77.7\%. The study also demonstrates that post-mergers are more difficult to classify than pre-mergers.
\subsection{Occlusion}

\begin{figure*}[htbp]
   \centering
   \includegraphics[width=\linewidth]{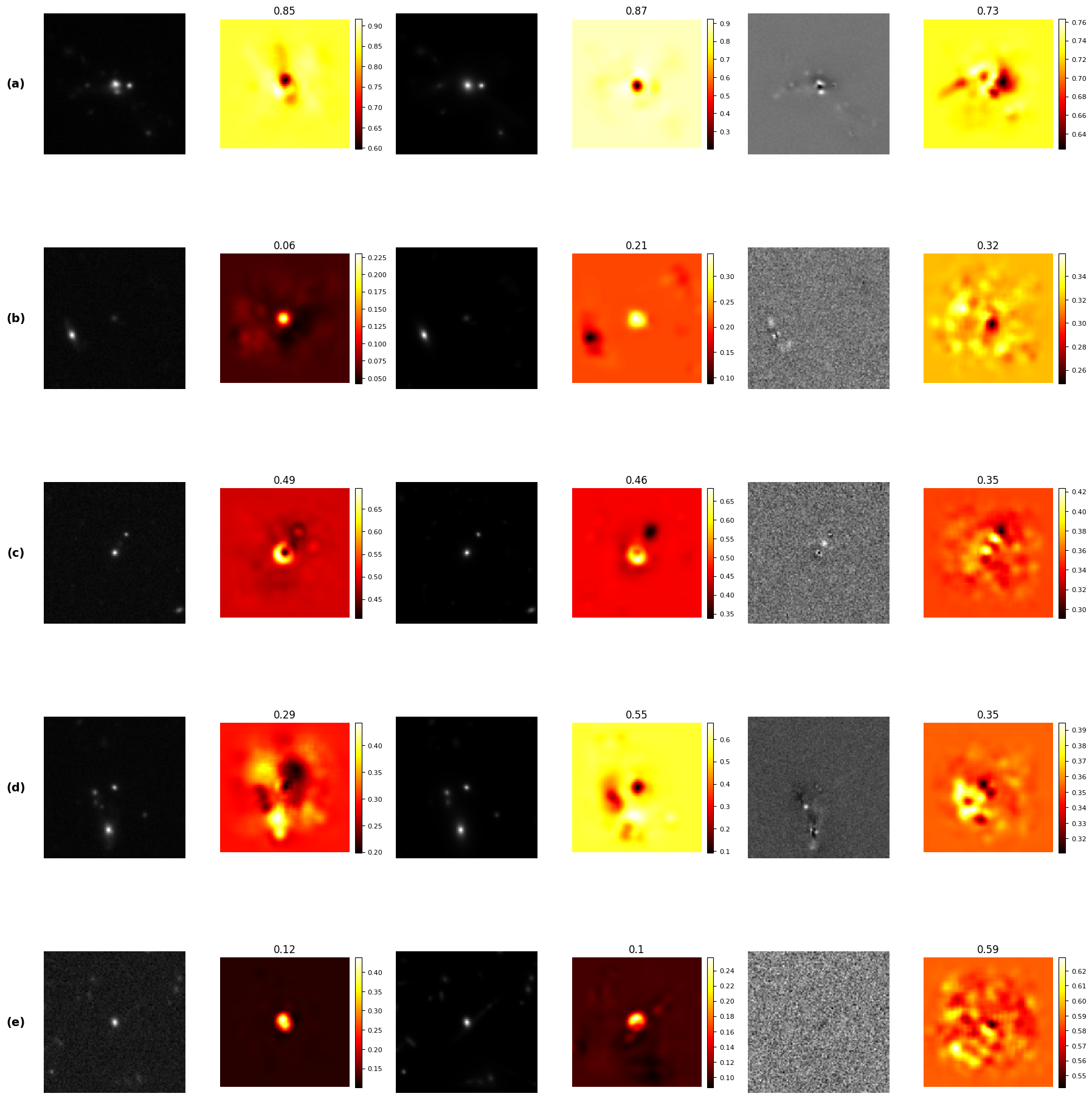}
      \caption{Occlusion experiment for mergers that were: a) classified as mergers by all three CNNs, b) classified as non-mergers by all three CNNs, c) classified as mergers by CNN1 and as non-mergers by CNN2 and CNN3, d) classified as mergers by CNN2, and as non-mergers by CNN1 and CNN3, and e) classified as mergers by CNN3 and as non-mergers by CNN1 and CNN2.  First two left columns show CNN1 input and corresponding heatmap. Two middle columns show CNN2 input and corresponding heatmap. Last  middle columns show CNN3 input and corresponding heatmap. Above each heatmap there is a prediction for the original input, before performing occlusion experiment. Thresholds for CNN1, CNN2, and CNN3 are 0.39, 0.47, and 0.49 respectively. Pixels where occlusion decreases the predicted merger probability are shown in darker red and correspond to regions of high feature importance, whereas pixels where occlusion increases the merger probability appear in lighter colors, indicating low importance for the classification}.
         \label{mergers_oc}
   \end{figure*}

\begin{figure*}[htbp]
   \centering
   \includegraphics[width=\linewidth]{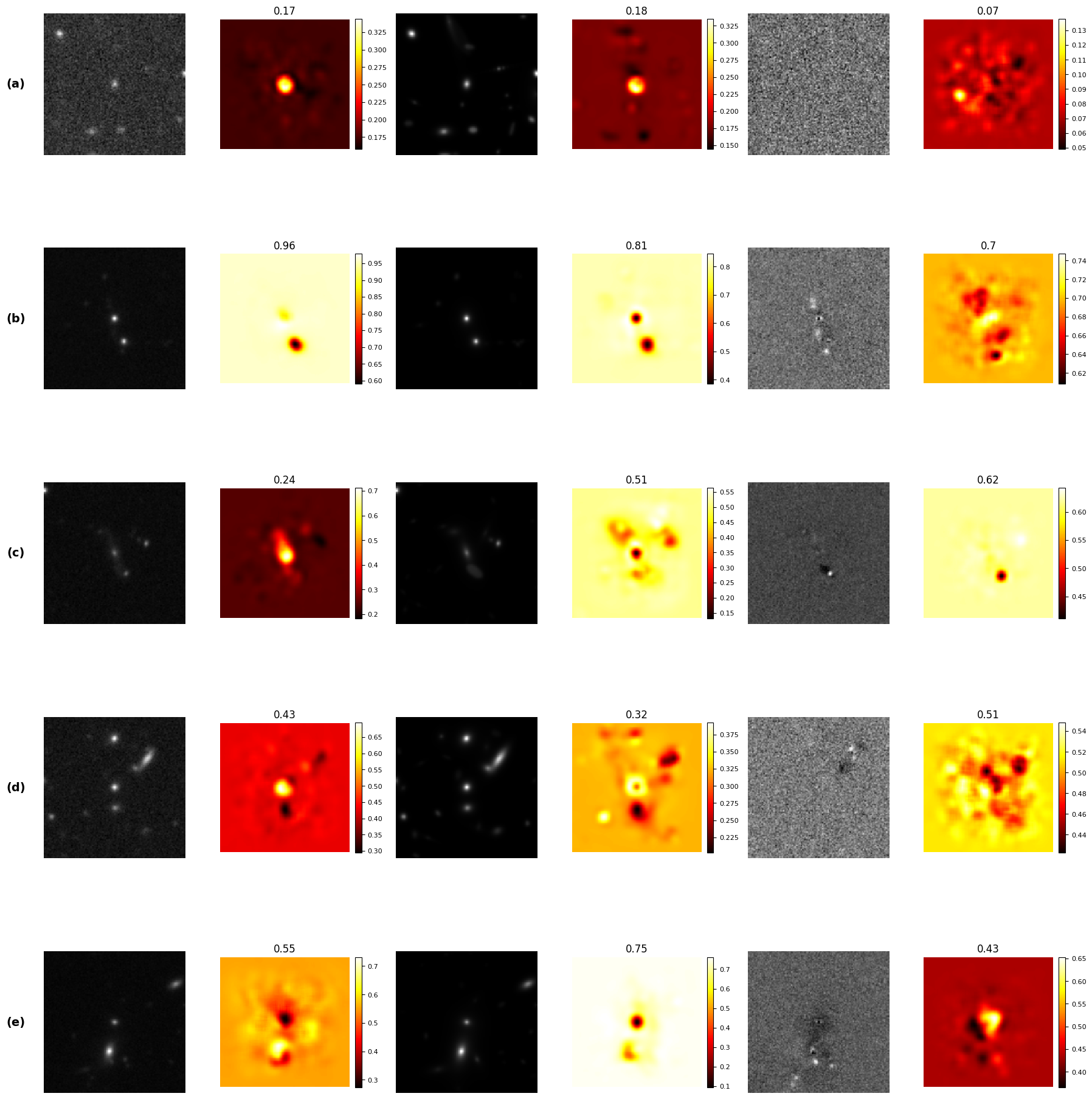}
      \caption{Occlusion experiment for non-mergers that were:  a) classified as non-mergers by all three CNNs, b) classified as mergers by all three CNNs, c) classified as non-mergers by CNN1 and as mergers by CNN2 and CNN3, d) classified as non-mergers by CNN2, and as mergers by CNN1 and CNN3, and e) classified as non-mergers by CNN3 and as mergers by CNN1 and CNN2. First two left columns show CNN1 input and corresponding heatmap. Two middle columns show CNN2 input and corresponding heatmap. Last two  columns show CNN3 input and corresponding heatmap. Above each heatmap there is a prediction for the original input, before performing occlusion experminet. Thresholds for CNN1, CNN2, and CNN3 are 0.39, 0.47, and 0.49 respectively. Pixels where occlusion increases the predicted non-merger probability are shown in darker red and correspond to regions of high feature importance, whereas pixels where occlusion decreases the merger probability appear in lighter colors, indicating low importance for the classification.}
         \label{non-mergers_oc}
   \end{figure*}

Following \cite{pearson2022north}, occlusion experiments were performed on five mergers and five non-mergers from the test set and the $0.76 < z < 1$ redshift range. This technique is used to visualize what parts of the image are important for classification, as CNN feature maps are very unintuitive in deeper layers.
The occlusion experiment relies on covering the image with a box (we used 8×8 pixels), evaluating the models' predicted class probability after occluding, shifting the box one pixel to the right so it covers different pixels, evaluating the class again, and repeating it all over until the box reaches the lower right corner of an image. The occluded images are treated as normal galaxy images by a network. For CNN1 and CNN3 we used a box filled with noise with the properties of the image background, and for CNN2 we used a box filled with 0 values. This is done to not introduce structures that were not previously seen by the network, as there are no empty chunks in original images and residual images and there are no noise structures on the model images. This technique allows us to create an importance heatmap, which represents the average classification when each pixel was occluded. For mergers, important parts of the image will have lower scores in the importance heatmap than less important parts, and for non-mergers, important parts of the image will have higher scores than less important parts.
In Fig. \ref{mergers_oc} we choose mergers that were: a) classified as mergers by all three CNNs, b) classified as non-mergers by all three CNNs, c) classified as mergers by CNN1 and as non-mergers by CNN2 and CNN3, d) classified as mergers by CNN2, and as non-mergers by CNN1 and CNN3, and e) classified as mergers by CNN3 and as non-mergers by CNN1 and CNN2. We also take non-mergers in Fig. \ref{non-mergers_oc} that were: a) classified as non-mergers by all three CNNs, b) classified as mergers by all three CNNs, c) classified as non-mergers by CNN1 and as mergers by CNN2 and CNN3, d) classified as non-mergers by CNN2, and as mergers by CNN1 and CNN3, and e) classified as non-mergers by CNN3 and as mergers by CNN1 and CNN2.

\begin{figure*}[htbp]
    \includegraphics[width=\linewidth]{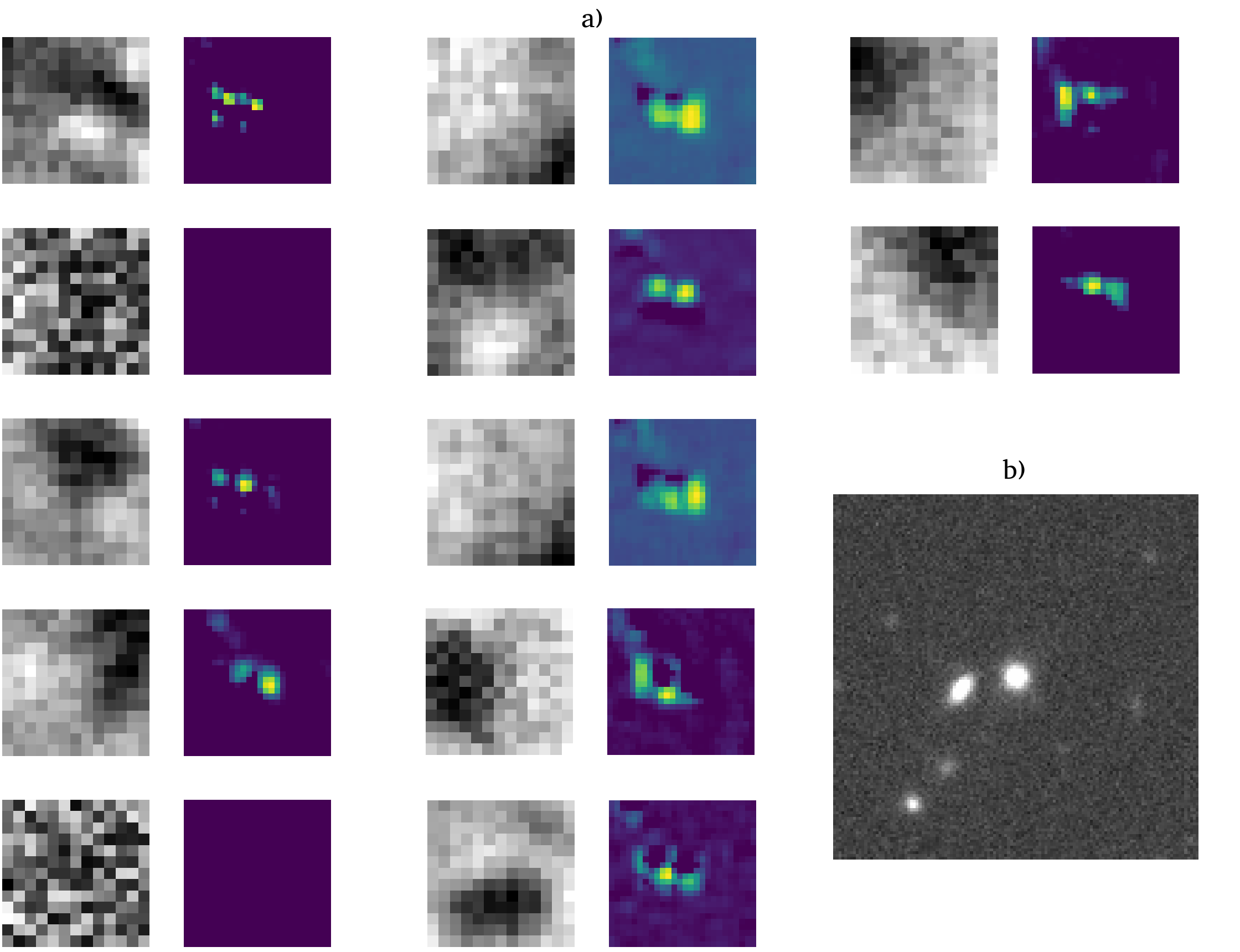}
\caption{Learnable kernels of CNN1 and their corresponding feature maps (a), along with the original merger image (b).}

         \label{fmaps}
   \end{figure*}

For selected known mergers (Fig.\ref{mergers_oc}), CNN1 chooses the primary galaxy as the most important part of the image for classification (a), (c) and (d), as seen by the lowest prediction values in the place where the primary galaxy is. It can also be seen that the secondary galaxy plays a role in the classification score in (b), though not that much as the outskirts of the primary galaxy. For selected known non-mergers (Fig. \ref{non-mergers_oc}) we observe similar behavior. CNN1 also uses the primary galaxy to determine the classification for (a), (b), (c), and (d), where in (e) we see a shift of focus to the secondary galaxy, which in this case may be due to its greater brightness than primary galaxy. As expected, the most important region for classification of raw images seems to be the primary galaxy. This is the region where double nuclei are expected to be observed in merging galaxies, which are strong indicators of a merging process. This can also be seen in feature maps - convolutions of image with learnable kernel in Fig. \ref{fmaps}, which confirms that double nuclei is probably one of the most important features for CNN1.

For selected known mergers (Fig.\ref{mergers_oc}) CNN2 is less influenced by the center galaxy than CNN1. The primary galaxy is the most important region for classification in (a), and (d), where in (b), (c), and (e) other galaxy models also contribute to the change of the score. For known non-mergers (Fig.\ref{non-mergers_oc}) CNN2 focuses on the primary galaxy only in (a). In (b), and in (d) it focuses on the outskirts of the primary galaxy. In (c) when it sees the secondary galaxy it changes its score in favor of a merger, and in (e) it focuses on the outskirts of both primary and secondary galaxy. For model images, the central region of the image also seems to be of the biggest importance for classification. Probably because the primary galaxy usually is the brightest in the image and the model network does not learn any additional information, just the shape and intensity of the model. In Fig. \ref{mergers_oc} (c) and in Fig. \ref{non-mergers_oc} (c) we see diminishing importance of the primary galaxy in favor of the secondary galaxy. This fact seems to support this idea, as it occurs when the size and intensity of both models are similar, and there is no single clearly dominating galaxy.

 In the case of non-mergers (Fig.\ref{non-mergers_oc}) CNN3 relies on the residual features more than for the merger case. This would agree with \cite{la2024dust} who looked at the activation maps and for the non-mergers the most relevant parts of the images were also the features surrounding the primary galaxy. We can see that residual features play a role in evaluating the score on  (b), (c), and (d), and (e).
For known mergers (Fig.\ref{mergers_oc}) CNN3 looks into structures around the center, where the primary galaxy was located in (b), (d), and (e). In (a) and (c) it focuses on place where the secondary galaxy used to be.  The residual network, as expected, focuses around the place where the primary galaxy used to be or its close surrounding. This has two explanations. Firstly - there can still be double nuclei features present as it may not follow the S{\'e}rsic profile and remains in the image, which is probably what happened in Fig. \ref{mergers_oc} (a) and (c) and Fig. \ref{non-mergers_oc} (b). Secondly - this is the region where faint merging features are expected to be present, and this can arguably be seen in Fig. \ref{mergers_oc} (e) and Fig. \ref{non-mergers_oc} (b), (d), and (e).

It is important to acknowledge that these occlusion experiments were conducted on a relatively small sample due to computational constraints. As a result, while the observed patterns are compelling, they may not fully capture the diversity of behaviours across the entire dataset. Nevertheless, the consistency in CNN focus areas, particularly around primary galaxies and potential double nuclei, provides promising evidence that the networks are learning meaningful astrophysical features. We are therefore optimistic that this is indeed what the networks are attending to when making their classifications.

Our occlusion results show strong qualitative consistency with the image-only network in \cite{pearson2022north}. In both cases the primary galaxy is the dominant feature driving the classification, with the lowest occlusion scores typically occurring where the central light concentration or potential double nuclei reside. Similar to \cite{pearson2022north}, our networks also show that bright secondary galaxies can influence the classification, sometimes being interpreted as potential companions even when they are unrelated. A notable difference is that our CNNs, especially the residual-based one, show a clearer sensitivity to faint structures around the primary galaxy, while \cite{pearson2022north} reported weaker evidence for such features in their occlusion maps. Overall, however, the importance patterns are broadly aligned, indicating that both sets of networks rely on comparable visual cues when identifying mergers.

Generating an occlusion map took 3 minutes per image on an RTX 4090.

\section{Conclusions} \label{conclusions}
We trained three CNNs to perform merger vs non-merger classification using the IllustrisTNG simulation data with mock images that mimic the HSC survey: CNN1 on raw images, CNN2 on S{\'e}rsic models, and CNN3 on residual images. Then we evaluated the performance of all CNNs on the test data. CNN1 achieved the best accuracy of 74\%. 

It is possible to classify merging and non-merging galaxies using source position and shape information contained in S{\'e}rsic profiles with 70\% accuracy (particularly pre-mergers with 77\% precision), and using residual images with faint structures present in the image, achieving 68\% accuracy (particularly post-mergers with 56\% precision). Studying the results we draw three important conclusions about the relative importance between spacial information and residual features for non-merger vs merger classification:
\begin{itemize}
    \item[\textbf{I.}]\textbf{Both source position, shape information and residual features are crucial for non-merger vs merger classification.}
    \begin{itemize}
        \item  The almost equal overall performance of CNN2 and CNN3 indicates that both spatial and residual information are important for distinguishing between mergers and non-mergers. The lack of either type of information reduces the network's performance, as CNN1, which has access to all types of information, performs best in this task.
        \item  The very good ability to recognize mergers at low ellipticities shows that, in addition to the source's position, its shape is also important for the network.
    \end{itemize}
    \item[\textbf{II.}]\textbf{Source position and shape information is crucial for pre-merger classification}
    \begin{itemize}
        \item Strong CNN1 and CNN2 performance compared to CNN3 in correctly classifying pre-mergers implies that source position and shape information is crucial for this task. The impact of its absence is observed in CNN3, which, lacking in relative position cues, performs significantly worse. Lack of a secondary galaxy near the primary galaxy strongly indicates a non-pre-merger event.
    \end{itemize}
    \item[\textbf{III.}]\textbf{Residual features are crucial for post-merger classification.}
    \begin{itemize}
        \item CNN2, devoid of residual context, performs the worst in classifying post-mergers, only 3\% better than random pick. CNN1 having all information correctly predicts 59\% of post-mergers. Introduction of residual features by subtracting S{\'e}rsic profiles drops the accuracy  by 3\% to 56\% as seen in CNN3.
        \item The quality of the image can affect the network's ability to classify post-mergers. Higher-quality images may have more visible merging features in the residuals.
    \end{itemize}
\end{itemize}    
Future work will focus on evaluating networks on real HSC images labeled by the Galaxy Zoo project \citep{lintott2011galaxy}. We will also apply the same data preprocessing method to the observations and train the networks. The differences between the performance of the simulation-trained network and the observations-trained network will be examined. New architectures will be tested to find the best way to learn both residual and spatial features.
\newpage
\section*{Acknowledgments}
We gratefully acknowledge Polish high-performance computing infrastructure PLGrid (HPC Center: ACK Cyfronet AGH) for providing computer facilities and support within computational grant no. PLG/2024/017417. We also thank Tomasz Łaguz for providing access to an RTX 4090, which enabled on-demand training, testing, and evaluation. W.J.P has been supported by the Polish National Science Center project UMO-2023/51/D/ST9/00147. A.P has been supported by the Polish National Science Center project UMO-2023/50/A/ST9/00579. L.E.S was supported by the Estonian Ministry of Education and Research (grant TK202), Estonian Research Council grant (PRG1006), and the European Union's Horizon Europe research and innovation programme (EXCOSM, grant No. 101159513). B.M.B and L.W are supported by the project ‘Clash of the Titans: deciphering the enigmatic role of cosmic collisions’ (with project number VI.Vidi.193.113 of the research programme Vidi, which is (partly) financed by the Dutch Research Council (NWO).
    
\appendix
 \begin{table}[htbp]
\centering
\caption{The CNN architecture.}
\begin{tabular}{c} 
\hline
Normalized input images (raw, model, residual), range [0,1] \\
\hline
Conv2D: 128 filters, 13×13 kernel \\
ReLU, Batch Normalization, Dropout (20\%), MaxPooling 2×2 \\
\hline
Conv2D: 128 filters, 11×11 kernel \\
ReLU, Batch Normalization, Dropout (20\%), MaxPooling 2×2 \\
\hline
Conv2D: 128 filters, 9×9 kernel \\
ReLU, Batch Normalization, Dropout (20\%), MaxPooling 2×2 \\
\hline
Flatten \\
\hline
FC1: 512 neurons \\
ReLU, Batch Normalization, Dropout (20\%) \\
\hline
FC2: 128 neurons \\
ReLU, Batch Normalization, Dropout (20\%) \\
\hline
Output layer: 1 neuron \\
Sigmoid \\
\hline
Loss: Binary cross-entropy \\
Optimizer: Adam ($\alpha = 3 \cdot 10^{-4}$) \\
Epochs: 50 \\
8\,966\,273 trainable parameters \\
\hline
\end{tabular}
\label{architecture_table}
\end{table}

\bibliography{mybibliography}


\end{document}